\documentclass[onecolumn,11pt]{article}
\usepackage[top=1in, bottom=1in, left=1in, right=1in]{geometry}

\usepackage{times}
\usepackage{sidecap}

\usepackage{dcolumn}
\usepackage{bm}
\usepackage{amsmath}
\usepackage{amssymb}
\usepackage{revsymb}
\usepackage{graphics}
\usepackage{graphicx}
\usepackage{setspace}
\usepackage{color}
\usepackage{wrapfig}

\usepackage{float}
\usepackage{enumitem}

\usepackage[font=small]{caption}

\usepackage{fancyhdr}
\begin{document}

\newpage
\centerline{\textbf{\LARGE{Harvesting Energy from Sun, Outer Space, and Soil}}}

\centerline{\textbf{Yanpei Tian$^1$, Xiaojie Liu$^1$, Fangqi Chen$^1$, Yi Zheng$^{1,2*}$}}

\centerline{$^1$Department of Mechanical and Industrial Engineering, Northeastern University, Boston, MA,USA.}
\centerline{$^2$Department of Electrical and Computer Engineering, Northeastern University, Boston, MA,USA.}
\centerline{$^*$e-mail: y.zheng@northeastern.edu}

\vspace{-10pt}
\section*{\large{Abstract}}
\vspace{-8pt}
While solar power systems have offered a wide variety of electricity generation approaches including photovoltaics, solar thermal power systems, and solar thermoelectric generators, the ability of generating electricity at both the daytime and nighttime with no necessity of energy storage still remains challenging. Here, we propose and verify a strategy of harvesting solar energy by solar heating during the daytime and harnessing the coldness of the outer space through radiative cooling to produce electricity at night using a commercial thermoelectric module. It enables electricity generation for 24 hours a day. We experimentally demonstrate a peak power density of 37 mW/m$^2$ at night and a peak value of 723 mW/m$^2$ during the daytime. A theoretical model that accurately predicts the performance of the device is developed and validated. The feature of 24-hour electricity generation shows great potential energy applications of off-grid and battery-free lighting and sensing. 

\newpage
\noindent
Energy crisis and environmental pollution have motivated the fundamental and applied investigations on a wide variety of renewable energy harvesting technologies \cite{cottrill2018ultra,reddy2013state,polman2012photonic,kraemer2011high}. While photovoltaics and solar thermal power systems are feasible approaches to generate electricity both for large-scale and off-grid applications over the daytime, they both rely on batteries or phase change materials to store electricity or heat at night \cite{hoppmann2014economic,xu2015application}, that drives up costs. Therefore, technologies of renewable power generation during both day and night with no necessity of storage are urgent for those people who lack reliable access to electricity in rural areas of the developing world \cite{cabraal2005productive}. From the thermodynamic perspective, there must be a hot reservoir and a cold sink to produce useful work as a heat engine for any kind of energy conversion process \cite{ondrechen1983generalized}. Solar thermal power systems use the Sun as a huge hot source and the surrounding environment as a cold sink to generate electricity. The Sun ($\sim$ 5800 K), the soil near the Earth's surface ($\sim$ 290 K) \cite{curiel2007microbial}, and the outer space ($\sim$ 3 K) are three separate locations with a huge temperature difference. Radiation heat transfer builds a bridge for these three locations through solar heating and radiative cooling. Solar heating: during the daytime, the ubiquitous and easily accessible sunlight serves as a heat source, and the chilly soil acts as a cold sink to build a temperature difference for the heat engine. Radiative cooling: at night, using the warm topsoil of the Earth instead as the heat reservoir and the outer space as the cold sink by radiating the heat out to the cold space through the atmospheric window (8 $\mu$m to 13 $\mu$m) to establish a reversible temperature difference \cite{raman2014passive}. Compared with the temperature of the air, that of the soil is less fluctuated, that is, the soil is colder than the air in the daytime, while warmer at night. Besides, the heat flux that pass through the soil and the thermoelectric generator (TEG) through heat conduction is higher than the natural convection when the cold side of TEG is cooled down by the surrounding by air. Therefore, the soil is an ideal alternative heat sink for the daytime and a heat source at night, which drives a 24-hour heat engine without energy storage technologies.

Solar heating relies on solar absorbers to convert sunlight into heat, and a unity absorptivity in the solar irradiance region (0.3 $\mu$m $\sim$ 2.5 $\mu$m) is preferred, while a nearly zero emissivity in the mid-infrared region (2.5 $\mu$m $\sim$ 20 $\mu$m) is favored to depress the thermal loss from the spontaneous thermal radiation \cite{orel2005spectrally}. Extensive investigations on selective solar absorbers have been conducted to achieve high performance by exploring photonic crystals \cite{chou2014enabling,chou2014design}, metamaterials \cite{wang2014selective,wu2012metamaterial}, and cermet \cite{okuyama1979ni,cao2015high}. These solar absorbers own good spectrally selective properties with 90\% $<$ $\alpha_{solar}$ $<$ 98\% and 3\% $<$ $\epsilon_{IR}$ $<$ 10\% \cite{konttinen2003mechanically}. However, radiative cooling that passively cool down a sky-facing surface below ambient temperature by accessing the coldness of the outer space through the highly transparent atmospheric window (8 $\mu$m $\sim$ 13 $\mu$m) \cite{rephaeli2013ultrabroadband} requires the exposed surface has an ideal unity emissivity over the atmospheric window. Previous studies on radiative cooling have demonstrated different material options including photonic structures \cite{rephaeli2013ultrabroadband,raman2014passive}, metamaterials \cite{zhai2017scalable}, and polymer nanofiber and aerogel \cite{leroy2019high,wang2020scalable}. Therefore, an ideal surface with an unity absorptivity from 0.3 $\mu$m to 2.5 $\mu$m and a convertible emissivity (0 $\leftrightarrow$ 1, day $\leftrightarrow$ night) from 8 $\mu$m to 13 $\mu$m is demanded to obtain a relatively high temperature difference during both the daytime and nighttime. The materials with time-dependent reversible emissivity are hard to achieve cost-efficiently, so the all-black surface with an unity absorptivity over both the solar radiation region and mid-infrared range is an alternative to harvest heat from the Sun and coldness from the outer space simultaneously. Polymer-based black paint is a low-cost substitute for large-scale engineering applications.

One feasible path to convert a temperature difference into available electricity is to employ a TEG. The viable TEG modules have been developed and widely utilized in vehicles \cite{bell2008cooling}, wearable devices \cite{jung2017wearable}, and industrial waste-heat recovery systems \cite{kajihara2015study,aranguren2015experimental} because of their unique advantages such as no-pollution, small modular availability and non-mechanical vibration \cite{xie2010design,xie2010design}. This provides us a reliable approach to reasonably take advantage of the naturally-existed temperature difference between the Sun, the Earth's topsoil, and the outer space. Here, we experimentally demonstrate an environment-friendly and cost-effective energy strategy for electricity generation both at the daytime and nighttime based on a TEG module. Such a system can produce a peak power density of 37 mW/m$^2$ at night which is higher than previously reported value (25 mW/m$^2$) \cite{raman2019generating} and a peak value of 723 mW/m$^2$ for the daytime without an energy storage system or no active power input. A analytical model is developed to characterize the system's performance under different configurations and weather conditions. The outdoor experiment demonstrates that the proposed device work 24 hours continuously and it becomes a great alternative for battery-free lighting and sensing in rural areas.

\section*{Results}

\subsection*{Optical Characterization}

\begin{figure}[!ht]
\centering
\includegraphics[width=0.85\textwidth]{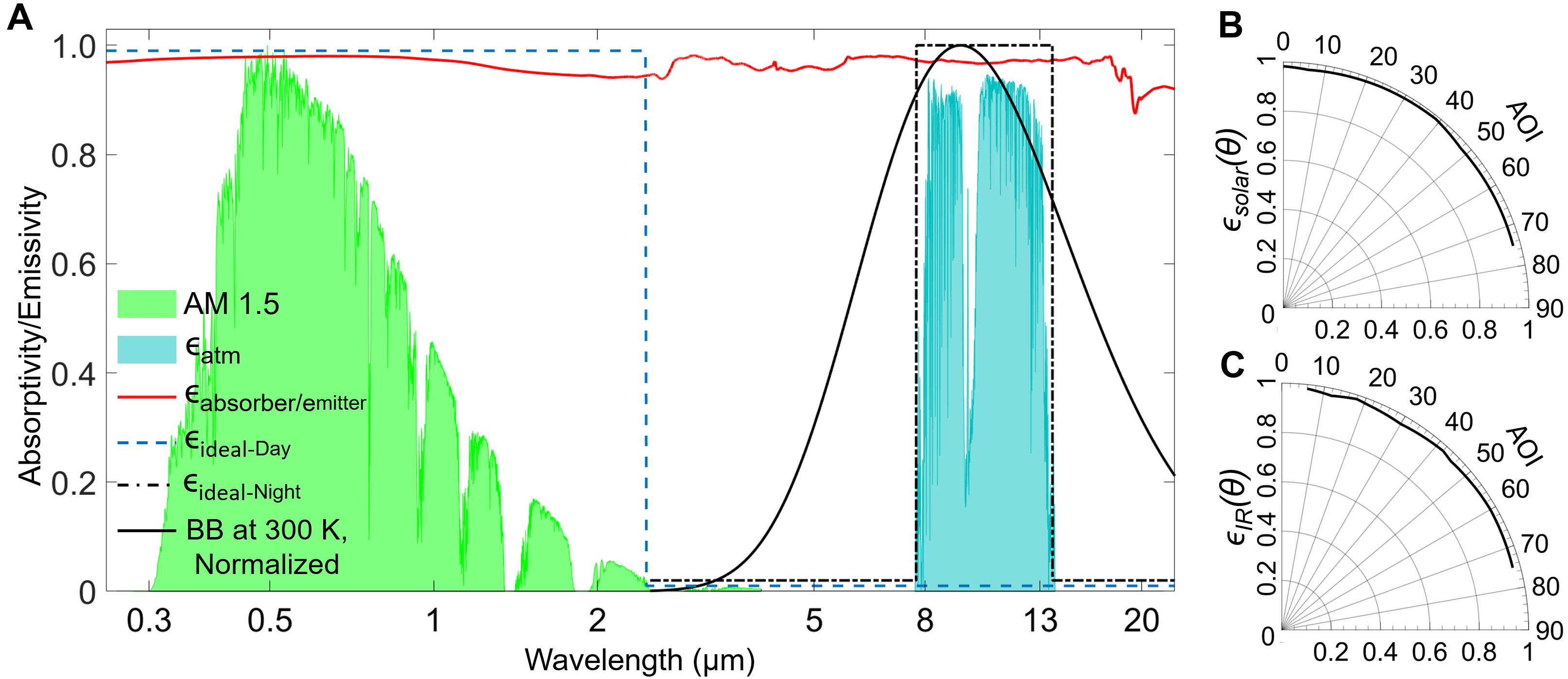}
\caption{\label{fig:spectra} \textbf{Optical properties of a black-painted copper plate.} (\textbf{A}) The emissivity spectra of the black 3.0 painted copper plate (0.62 mm thick) presenting with the normalized ASTM G173 Global solar spectrum, the mid-infrared atmospheric transparency window, and a normalized blackbody radiation spectrum at 300 K. The blue dashed line and black dotdashed line show the ideal spectra of a selective solar absorber for the daytime and radiative cooler for the nighttime, respectively. Excellent $\epsilon_{solar}$ ($\theta$) (\textbf{B})  and $\epsilon_{IR}$ ($\theta$) (\textbf{C}) across different angles of incident (AOI) result in  angle-independent, excellent hemispherical emissivity of the the black-painted copper surface.
} 
\end{figure}

\begin{figure}[!t]
\centering
\includegraphics[width=0.8\textwidth]{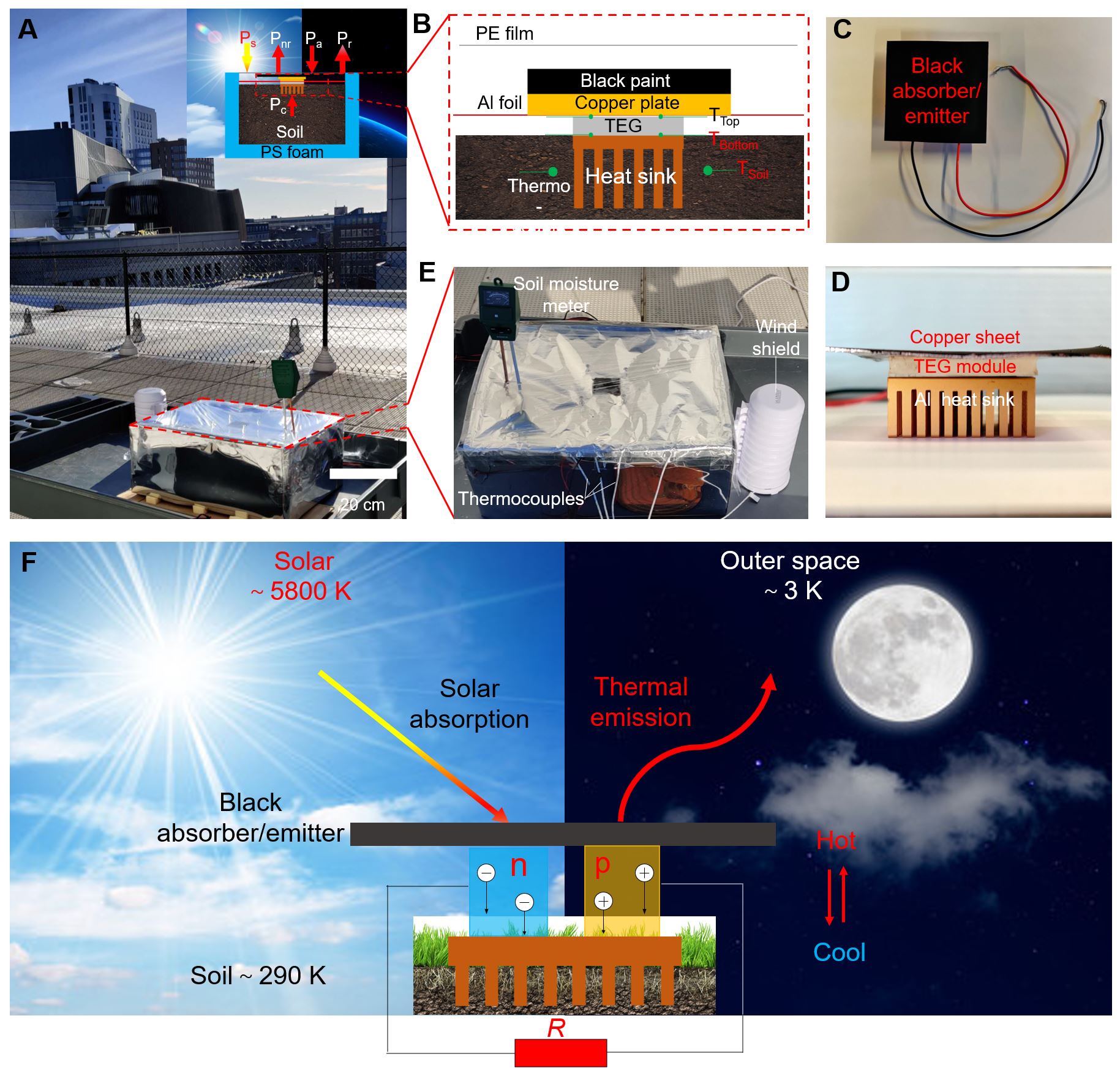}
\caption{\label{fig:setup} \textbf{Experimental investigation of a 24-hour TEG based energy harvesting system.} (\textbf{A}) Photography showing a 24-hour TEG device in rooftop testing at Northeastern University, Boston, MA. The inset depicts the schematic of the cross-section and the energy flows between TEG, the Sun, the ambient, and the outer space. (\textbf{B}) Schematic of the 24-hour TEG module consisting of the black-coated dual-purpose copper plate as both a solar heater and a radiative cooler. The TEG module is connected with the copper plate and heat sink using the thermal compound paste. An aluminum heat sink is inserted into the soil to release the heat during the daytime and absorb it at night. Top (\textbf{C}) and side (\textbf{D}) views of the TEG module.(\textbf{E}) Top view of the device with a soil humidity meter to show the moisture level in the soil, a windshield with a K-type thermocouple inside monitoring the ambient temperature, and thermocouples recording the temperature of the top and bottom surfaces of TEG module and two different locations in the soil. (\textbf{F}) Schematic of exhibiting the operational principle of a 24-hour TEG based energy harvesting system between the Sun, outer space, and soil. 
} 
\end{figure}

The spectra of an ideal selective solar absorber and a radiative cooler are shown in Fig \ref{fig:spectra}\textbf{A}. The selective solar absorber has a broadband spectral selectivity from 0.3 $\mu$m to 20 $\mu$m and shows a sharp cutoff at 2.5 $\mu$m. The ideal radiative cooler has a narrowband selective emissivity spectrum in the thermal radiation wavelength (2.5 $\mu$m to 20 $\mu$m) and has an abrupt rise and sudden drop at the edge wavelength of the atmospheric window (8 $\mu$m and 13$\mu$m, respectively). The stagnation temperature and net heating power of the ideal solar absorber exceeds the performance of the black emitter (Fig. S1\textbf{A} and S1\textbf{B}) at different weather conditions. The equilibrium temperature drops below ambient temperature and net radiative cooling power of an ideal radiative cooler also surpasses that of an infrared-black radiative cooler with different convection heat transfer coefficients. However, the revertible transformation of the emissivity spectra in the infrared region is hard to achieve without using the phase change materials, like VO$_2$ \cite{moore2017growth}. The visible- and infrared-black materials, e.g. black acrylic paint, stand out for its low cost and simple use on a wide variety of surfaces. Figure \ref{fig:spectra}\textbf{A} shows the spectral hemispherical emissivity of the black absorber/emitter. The nearly unity absorptivity ($\epsilon_{solar}$ = 0.98) in the solar irradiance region (0.3 $\mu$m $\sim$ 2.5 $\mu$m) ensures an excellent absorption of sunlight to heat up the top surface of the TEG module under the Sun. Additionally, the high $\epsilon_{solar}$ ($\theta$) from 0$^\circ$ to 75$^\circ$ incident angles secure that the solar absorber can efficiently absorb most of the sunlight, that eliminates the introduction of the energy-consuming solar tracker. Meanwhile, an excellent $\epsilon_{IR}$ ($\epsilon_{IR}$ = 0.974) of the radiative cooler can emit heat to the space to reduce its temperature, and a nearly unity $\epsilon_{solar}$ ($\theta$) from 6 $^\circ$ to 75$^\circ$ incident angles ensures the radiative cooler to release heat efficiently to the sky. 

\subsection*{Experimental Demonstration}

\begin{figure}[!ht]
\centering
\includegraphics[width=0.85\textwidth]{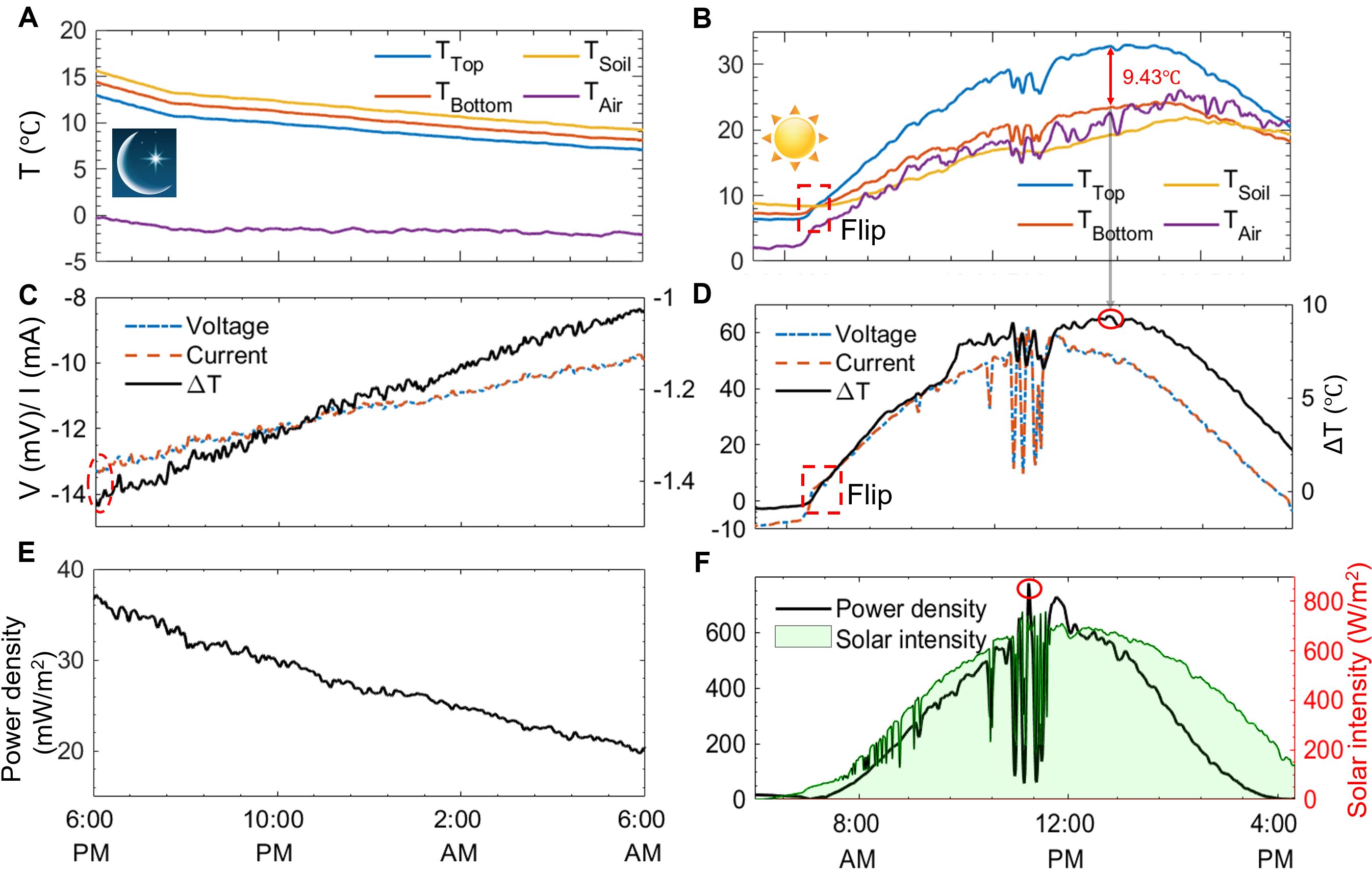}
\caption{\label{fig:data} \textbf{Passive working performance of the proposed 24-hour TEG system.} Temperature variations of the top and bottom surfaces of the TEG module, the soil, and ambient air at nighttime (\textbf{A}) and daytime (\textbf{B}) (the red dashed square circles the point where the temperature of top surface of the TEG module exceeds its bottom surface. The temperature difference between the top and bottom surfaces of the TEG module and its voltage and current output at night (\textbf{C})  and during the daytime (\textbf{D}) (the red solid oval marks the maximum point of $\Delta$$T_{Top-Bottom}$ of the TEG module). The output power density of the TEG module working in darkness (\textbf{E}) and operating under the sunlight (\textbf{F}). The right y-axis shows the variations of the solar intensity (the red solid oval marks the maximum point of the output power density).
} 
\end{figure}

To demonstrate the potential of the 24-hour TEG based power generation system utilizing the radiative energy transfer between the Sun, the soil, and the outer space, we built and tested a simple and low-cost TEG based energy harvesting system (less than \$15) whose top surface is coupled to a black absorber/emitter facing the sky and the bottom surface is connected to an aluminum heat sink that is inserted into the soil. The device is schematically shown in the inset of Fig. \ref{fig:setup}\textbf{A} and it consists of a all-black absorber/emitter, a TEG module and an aluminum heat sink (Fig. \ref{fig:setup}\textbf{B}). The thermal absorber/emitter is formed by a 70 mm $\times$ 70 mm $\times$ 0.8 mm copper sheet painted with a roughly 0.1 mm thick commercial black 3.0 paint ($\epsilon$ = 0.98) . The copper sheet adheres to the top surface of the commercial TEG module (SP1848-27145, ZT = 0.7) with a commercial thermal compound paste (ARCTIC MX-4, thermal conductivity: 8.5 W/(m$\cdot$K)). The bottom surface of the TEG module is coupled to an aluminum heat sink using thermal compound paste and the heat sink with aligned strip fins is immersed in the soil (Fig. \ref{fig:setup}\textbf{C} and \ref{fig:setup}\textbf{D} show the photos of the TEG device from the top and side views, respectively). The soil with the TEG device is packaged inside a 25 mm thick PVC insulation foam box covered by an aluminized mylar film to depress the thermal radiation from the box. A visible- and infrared-transparent convection shield made of 12.7 $\mu$m low-density polyethylene (LDPE) film to ensure the sunlight to go inside and the thermal infrared radiation to escape out (Fig. S2 displays the transmission spectrum of the LDPE film). The aluminum foil cover between the copper sheet and TEG module is for preventing the moisture from the soil condensing on the inner surface of the LDPE film, meanwhile blocking the sunlight and thermal infrared radiation to come in and out. The soil moisture meter is used to monitor the moisture level of the soil and keep it identical for different experiments (Fig. \ref{fig:setup}\textbf{D}). The outdoor experiment was conducted on a rooftop in Northeastern University (Boston, MA). The entire setup sits on a two-shelf utility cart which is about 0.8 m above the roof floor (Fig. \ref{fig:setup}\textbf{A}). Fig. \ref{fig:setup}\textbf{F} presents the basic operating principle of the proposed 24-hour TEG device, different from traditional TEG module, this device couples the hot side (top surface) to an ultra-black surface that converts sunlight to heat, while immerses and cools down its cold side (bottom surface) by the chilly soil at day. Conversely, the device combines its cold side (top surface) of the TEG module to a radiative-cooled surface that radiates heat to the outer space and enables its hot side (bottom surface) heated by the surrounding warm soil at night. Radiative energy exchange between the three different objects with huge temperature differences ($T_{Solar-Soil}$ $\approx$ 5510 K and $T_{Soil-Space}$ $\approx$ 287 K) ensures the TEG device working continuously for 24 hours. The hot/cold side of the TEG module is switched day and night.

We tested the performance of the TEG device at night in late January 2020 under a clear sky with an average ambient temperature of --1.6$^\circ$C and a dew point range from -- 15.3$^\circ$C  to -- 13.1$^\circ$C. The test started from 6:00 PM of January 29 after the sunset to 6:00 AM of January 30 before the sunrise. The soil temperature ($T_{Soil}$), the top surface temperature of the TEG module ($T_{Top}$), and bottom ($T_{Bottom}$) surface are monitored by K-type thermocouples connected to a National Instruments (NI) data acquisition board (PXI 6289). The thermocouple heads are located 5 cm near the aluminum heat sink to record the soil temperature. The thermocouple for recording the temperature of the ambient air is encapsulated inside a windshield to avoid frequent fluctuations due to the wind. The TEG module connects with an electric resistance ($R$ = 1 $\Omega$), and the output voltage and current are also recorded by NI PXI 6289. The top surface of the radiative cooler faces to the clear sky and draws heat from the topside of the TEG module to the cold outer space. The bottom side of the TEG adheres to the aluminum heat sink which receives the heat from the soil. Figure \ref{fig:data}\textbf{A} shows the temperature variations of the top side and bottom side of TEG module along with the measured ambient air and soil temperatures. The TEG's top surface, connected with the radiative cooler, is averaged to be 1.22$^\circ$C below its bottom surface and a temperature difference of up to 1.41$^\circ$C is observed from the recorded experimental data. The TEG's bottom surface is 1.12$^\circ$C lower than the soil. This temperature gradient from the soil to the cold space forms the energy flux through the TEG module. A maximum of 0.20 mW power is generated by the TEG module corresponding to the maximum power density of 37 mW/m$^2$ normalized to the area of the radiative cooler. The fluctuation of the power generation correlates with the temperature difference between the top surface and bottom surfaces since a bigger temperature difference represents a higher net radiative cooling power. The soil can be considered a big heat source to provide enough heat to the TEG module. The temperature of the soil is 15.37$^\circ$C higher than the ambient air and has fewer fluctuations than the air (Fig. \ref{fig:data}\textbf{A}), and the temperature difference  $\Delta$$T_{Soil-Bottom}$ is also smaller than $\Delta$$T_{Bottom-Air}$ (Fig. S3\textbf{A} and S3\textbf{B}).

The outdoor test of the TEG device for the daytime was conducted on March 4, 2020, with a peak solar intensity of 789 W/m$^2$ from 6:00 AM to 4:20 PM. The temperature difference between the top and bottom surface of TEG module flip at 7:20 AM, i.e. the hot side of the TEG switches from its bottom surface to its top surface (red dashed square in Fig. \ref{fig:data}\textbf{B} and \ref{fig:data}\textbf{D}), since the heating power from the absorption of solar irradiance neutralizes the cooling power due to radiative cooling. It is also reflected in the voltage curve in Fig. \ref{fig:data}\textbf{D}. The hot side of the TEG module turns from its bottom surface to the top surface after 7:20 AM. After that, the temperature of ambient air, the hot cold sides fluctuate with the variations of solar intensity, and it is particularly obvious from 10:30 AM to 11:40 PM during which the solar intensity fluctuates mightily. The maximum temperature difference of 9.43$^\circ$C between the hot and cold side occurs at 12:50 PM, at which the solar intensity is not at its maximum, because of the delay effect of the increment of the soil temperature. Besides, the soil temperature is below the ambient air most of the day, which proves that it is a better heat sink for cooling the TEG compared with the air (Fig. S3\textbf{C} and S3\textbf{D}). The maximum of the output power density occurs at the same time as the peak of the solar intensity (11:15 AM), which is inconsistent with the summit of the $\Delta$$T_{Top-Bottom}$ at 12:50 PM (Fig. \ref{fig:data}\textbf{D} and \ref{fig:data}\textbf{F}). It results from that the out power of the TEG module is determined by the heat energy absorbed by the solar absorber and then passes through the TEG module. The relative humidity (RH, \%) and wind speed (km/h) data for outdoor tests are presented in Fig. S4.

\section*{Discussions}
\begin{figure}[!ht]
\centering
\includegraphics[width=0.95\textwidth]{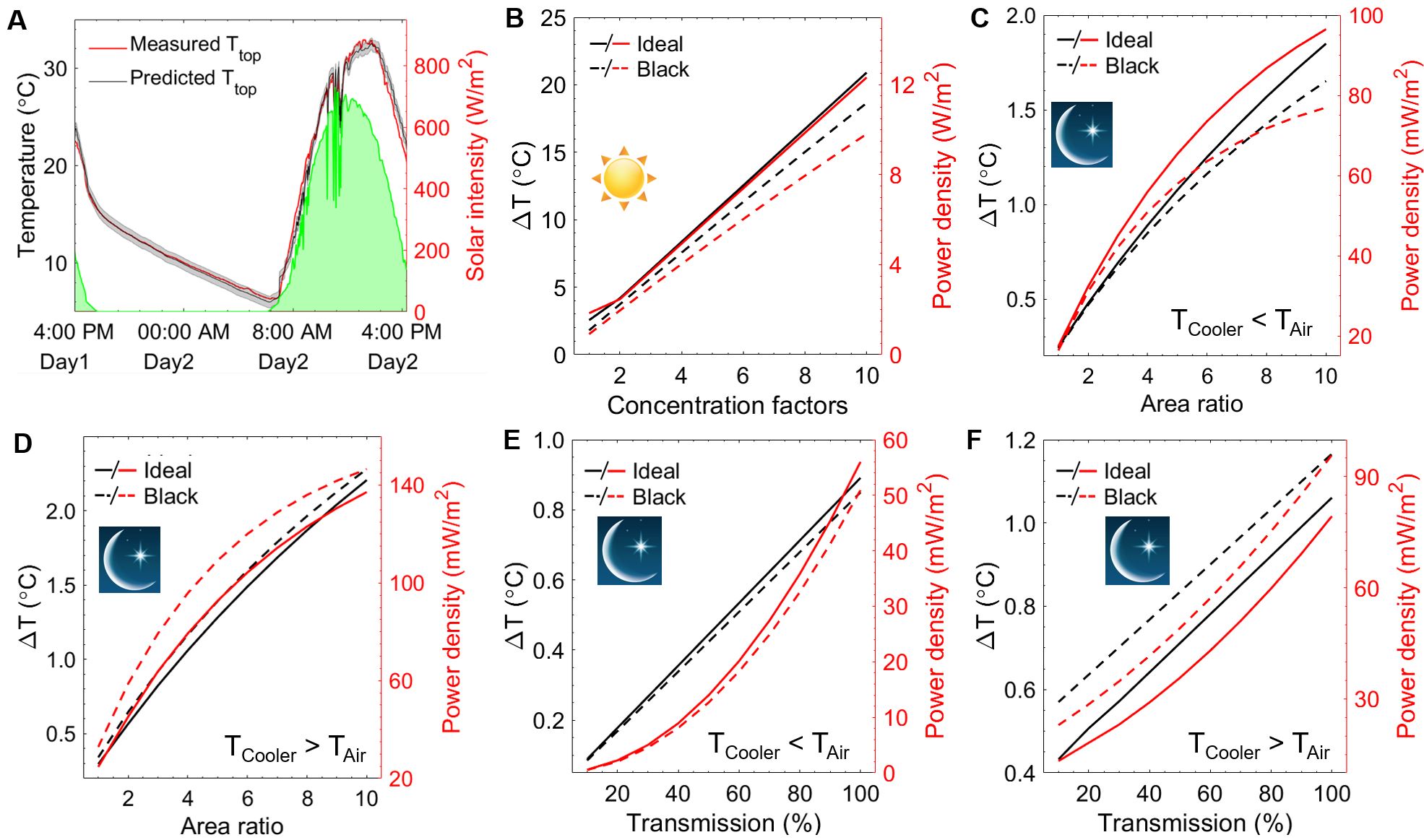}
\caption{\label{fig:final} \textbf{Theoretical model validation and extrapolation.} (\textbf{A}) Comparison between the top surface temperature of the TEG module predicted by the theoretical thermal model and experimentally measured one manifesting with the solar intensity. Model extrapolation: The temperature difference between the top and bottom surfaces of the TEG module and the output power density equipped with the ideal solar absorbers and the black ones as a function of solar concentration factors (\textbf{B}). The temperature difference achieved by radiative cooling of the top and bottom surfaces along with the output power density as a function of the area ratio (the radiative cooler surface over the TEG working surface) under different working environments ($T_{Cooler}$ $<$ $T_{Air}$ $\rightarrow$ (\textbf{C}); $T_{Cooler}$ $>$ $T_{Air}$ $\rightarrow$ (\textbf{D})) and different atmospheric transmittance ($T_{Cooler}$ $<$ $T_{Air}$ $\rightarrow$ (\textbf{E}); $T_{Cooler}$ $>$ $T_{Air}$ $\rightarrow$ (\textbf{F})). 
} 
\end{figure}

To evaluate the theoretical model's capability to track the experimental data (details of the theoretical model are provided in supplementary information), we take the measured solar intensity, ambient temperature and soil temperature as inputs to calculate $T_{{Black}}$. The non-radiative heat transfer coefficient, $h$, is evaluated in a range of 1 $\sim$ 3 W/(m$^2$$\cdot$K) considering that the LDPE convection shield has blocked most of the convective heat transfer. The thermal resistance between the aluminum heat sink and the soil, $R_{Sink-Soil}$, is limited within 0.0075 $\sim$ 0.01 m$^2$$\cdot$K/W. After determining the transient temperature of the absorber top surface, a gray band confines the bounds of the model's predictions considering the ranges of $h$ and $R_{Sink-Soil}$ (Fig. \ref{fig:final} \textbf{A}). The mean bias error (MBE) and the root mean square error (RMSE) for the nighttime are 0.08$^\circ$C and 0.37$^\circ$C, respectively, while MBE = 0.043$^\circ$C and RMSE = 1.22$^\circ$C during the daytime. It shows that the predicted values of the model are in good agreement with the experimentally measured ones.

With the validated thermal model, we can employ it to predict the temperature difference of the hot and cold sides for the TEG module in different working conditions, i.e. concentration factors of the sunlight, different configurations, area ratio of the solar absorber relative to the TEG surface, and different weather conditions. Figure. \ref{fig:final}\textbf{B} shows the temperature difference of TEG  hot and cold sides and the output power density at different sunlight concentration factors (the solar absorber area is the same of the TEG working surface's). Here, it is assumed that the ambient temperature is 25$^\circ$C and the soil temperature is 15$^\circ$C acting as a big heat sink, and the lossless non-radiative heat transfer ($h$ = 0). The sky is clear and The ZT (figure of merit) of the TEG does not vary during the day. With the increasing of concentration factors from 1 to 10, the temperature difference of hot and cold sides gets 10 times bigger when the TEG is equipped with an ideal solar absorber (Fig. \ref{fig:spectra}\textbf{A} blue dashed line), while the temperature difference only increases 4.9 times when the TEG is equipped with an actual black absorber. It proves the advantages of the selective solar absorber in the solar thermal related applications. The output power density is given by $W_{\max }=n \alpha^{2}\left(T_{h}-T_{c}\right)^{2} / 4 R A$, where $n$ = 127 is the number of the thermocouples pairs in the TEG module, $\alpha$ = 190 $\mu$V/K is the Seebeck coefficient, $R$ = 0.0065 $\Omega$ is the resistance of each thermocouple pair, and $A$ = 4.9 $\times$ 10$^{-3}$ m$^2$ is the area of the solar absorber. The power density changes almost linearly with the varying temperature difference, since the solar and the soil are assumed to be the big heat source and sink, respectively.  It yields that the power density increases as more energy flow through the TEG module with an increasing concentrated factors considering a constant efficiency of the TEG module.

The temperature difference of the TEG module and the power density as a function of the area ratio of the area of the radiative cooler relative to that of the TEG working surface are shown in Fig. \ref{fig:final}\textbf{C} and \ref{fig:final}\textbf{D}. A maximum power density of 147 W/m$^2$ can be achieved when the temperature difference is 2.3$^\circ$C with an area ratio of 10. In comparison between Fig. \ref{fig:final}\textbf{C} and \ref{fig:final}\textbf{D}, it can be found that the black emitter is more advantageous than an ideal one when $T_{Cooler}$ $>$ $T_{Air}$. Here, the ambient act as a heat sink, like space, for the radiative cooler to release heat since the air is highly absorptive from 5 $\mu$m to 8 $\mu$m and from 13 $\mu$m to 16 $\mu$m. However, the ideal radiative cooler has a better performance than the actual black one when $T_{Cooler}$ $<$ $T_{Air}$, since the nearly zero absorptivity prevents it from absorbing the heat from absorbing the heat from the ambient air through radiative heat transfer. The atmospheric transmittance related to the weather condition (relative humidity and cloud thickness) also plays a key role in the radiative cooling performance of the cooler. The power density of the TEG module drops as the averaged transmittance decrease, e.g. the power density reduces 75\% when the transmittance decreases from 100\% at an ideal clear sky condition to 50\% (Fig. \ref{fig:final}\textbf{E}). The same rules can be applied to the situations when $T_{Cooler}$ is higher than $T_{Air}$. The details on the temperature of the top and bottom sides for TEG module in Fig. \ref{fig:final}\textbf{B} -- \ref{fig:final}\textbf{F} are provided in Fig. S5.

This work experimentally validates the possibility of a TEG based energy harvesting system working 24 hour continuously through solar heating and radiative cooling by effectively assembling the commercially available and cost-effective TEG components. The prototype of the designed TEG device is prepared by using a high efficient black painted absorber/emitter to absorb the solar energy and harvesting the coldness of the outer space. The soil acts perfectly as a heat sink at daytime, while as a heat source for the nighttime, that resulting in a better performance than the ambient air. The theoretical model reveals that the concentration factor and the area ratio are key 
design parameters to achieve a maximum efficiency. The weather conditions, e.g. relative humidity and cloud thickness, are studied to evaluate the performance of the proposed device. It shows that selective radiative cooler is not always a better option than the black one which is easier to fabricate using the commercial black paints. This approach provides an alternative for the electricity generation at both daytime and nighttime without the necessity of the use of the battery storage for the off-grid rural areas. The Sun, the soil, and the outer space are ubiquitous for everyone, that renders it unlimited and unconstrained for the future energy applications.

\section*{Materials}
The black 3.0 paint is purchased from Culture Hustle USA and used without purification. The 0.8 mm thick copper sheet (152 mm $\times$ 152 mm) is from Integrity Beads and then cut into pieces on demand. The thermoelectric generator module (SP1848-27145 SA, 40 mm $\times$ 40 mm $\times$ 3.4 mm) is provided by Akozon. The aluminum heatsink (37.6 mm $\times$ 36.6 mm $\times$ 23.6 mm) is from Electronic-Salon. The thermal compound paste is from Arctic. K-thermocouple temperatures (measurement range, -- 40$^\circ$C $\sim$ 350$^\circ$C) are provided by Twtade. Miracle Gro garden soil is purchased from the Home Depot. The silver mylar film is provided by Lepilion. The LDPE plastic film is gotten from Glad Cling. The soil moisture meter is purchased from Sonkir.

\section*{Methods}
\subsection*{Preparation of the Black Absorber/emitter}
The 0.8mm thick copper sheet is cut into 70mm $\times$ 70 mm size and ultrasonic-washed by the acetone and DI water to remove the surface oil, and then blow-dried with the compression Argon. The 3 ml black 3.0 paint is thinned with 1.2 ml DI water under vigorous stirring for 5 min with a paintbrush to get a homogeneous mixture. Subsequently, the mixed black 3.0 paint is sprayed onto the copper sheet by a touch-up spray gun (Paasche Airbrush, USA) with a 0.8 mm spray head at a pressure of 70 psi. The distance between the spray head and the copper is kept at about 25 cm. The spraying process is repeated by 4 times to get a full coverage of the copper surface with a uniform 0.01mm thick black 3.0 paint layer and then the copper sheet is dried with a hot air blower (Yihua Electronic Equipment Co., Ltd, Guangzhou, China) at a temperature of 190$^\circ$C for 5 min. The distance between the copper and the hot air blower head is about 10 cm.

\subsection*{Emissivity Spectra Measurement}
The reflectivity spectra (UV-Visible-Near-infrared range: 200 nm $\sim$ 2500 nm) are measured by the Jasco V770 spectrophotometer at an incident angle of 6$^\circ$ with the ISN-923 60 mm BaSO$_4$ based integrating sphere equipped with PMT and PbS detectors. The reflectivity spectra are normalized by a PTFE based reflectance standard. The reflectivity spectra (Mid-infrared range: 2.5 $\mu$m $\sim$ 20 $\mu$m) is measured by Jasco FTIR 6600 at an incident angle of 12$^\circ$ with the PIKE upward gold integrating sphere equipped with wide-band MCT detector.  

\subsection*{Date Recording}
The temperature, current, and voltage are measured using the K-type thermocouples connected to the National Instruments PXI-6289 multifunction I/O module. The relative humidity (RH, \%), wind speed (km/h), and solar intensity (W/m$^2$) are measured by Ambient Weather WS-2000 smart weather station.

\section*{Data availability}
The datasets analyzed during the current study are available
within the paper, its Supplementary Information.  All other data related to this work are available from the corresponding author upon reasonable request.

\newpage
\bibliography{Yanpei}

\newpage
\section*{Acknowledgments}
This project is supported by the National Science Foundation through grant number CBET-1941743.

\section*{Author contributions}
Y.T., X.L., and Y.Z. develop this concept. Y.T. develops the theoretical model and writes the manuscript with help from all other authors. Y.T. and X.L. conduct experimental investigations. All authors provide critical feedback and help revise the final version of the manuscript. Y.Z. supervises this project.

\section*{Competing interests}
The authors declare no conflict of interest.

\end{document}